\documentclass[aps,pre,showpacs,twocolumn,superscriptaddress]{revtex4-1}
\usepackage{graphicx}
\usepackage{amsmath, amssymb}
\begin{document}
\title{Fidelity decay for local perturbations: Microwave evidence for oscillating decay exponents}

\author{Bernd~K\"{o}ber}
\affiliation{Fachbereich Physik der Philipps-Universit\"at Marburg, D-35032
Marburg, Germany}

\author{Ulrich~Kuhl}
\affiliation{Fachbereich Physik der Philipps-Universit\"at Marburg, D-35032
Marburg, Germany}
\affiliation{Laboratoire de Physique de la Mati\`{e}re Condens\'{e}e, CNRS UMR 6622, Universit\'{e} de Nice Sophia-Antipolis, F-06108 Nice, France}

\author{Hans-J\"{u}rgen~St\"{o}ckmann}
\affiliation{Fachbereich Physik der Philipps-Universit\"at Marburg, D-35032
Marburg, Germany}

\author{Arseni~Goussev}
\affiliation{School of Mathematics, University of Bristol, University Walk, Bristol BS8 1TW, United Kingdom}

\author{Klaus~Richter}
\affiliation{Institut f\"{u}r Theoretische Physik, Universit\"at Regensburg, D-93040 Regensburg, Germany}

\date{\today}

\begin{abstract}
 We study fidelity decay in classically chaotic microwave billiards
 for a local, pistonlike boundary perturbation. We experimentally
 verify a predicted nonmonotonic crossover from the Fermi golden
 rule to the escape-rate regime of the Loschmidt echo decay with
 increasing local boundary perturbation. In particular, we observe
 pronounced oscillations of the decay rate as a function of the
 piston position which quantitatively agree with corresponding
 theoretical results based on a refined semiclassical approach for
 local boundary perturbations.
\end{abstract}

\pacs{05.45.Mt, 03.65.Sq}
\keywords{fidelity, scattering theory}

\maketitle

\section{Introduction}
\label{sec:Intro}

The stability of quantum time evolution measured by the overlap between time-evolved perturbed and unperturbed states,
as suggested by Peres \cite{per84}, has been studied from various viewpoints and under different names.
In the field of quantum information this overlap is called ``fidelity'' \cite{nie00} and plays an important
role for quantifying the susceptibility of quantum dynamics to environmental or other external perturbations.
In semiclassical quantum and wave mechanics, alternatively, the overlap of an initial state with the state reached
after successive forward and backward time propagation, governed by the unperturbed and perturbed Hamiltonian,
is often termed ``Loschmidt echo'' (LE) \cite{Usa98}, especially for Hamiltonians associated with
complex, in particular, classically chaotic dynamics.
This terminology refers to the notion of echoes from momenta reversal in a Hamiltonian system
considered by Loschmidt \cite{los76} in the 19th century.

For chaotic systems the LE has been predicted to exhibit different
decay characteristics \cite{Jal01} depending on the form and
strength of the perturbation. One distinguishes roughly three
prominent LE decay regimes: the perturbative Gaussian
\cite{Jac01b,Cer02}, the Fermi-golden-rule (FGR)
\cite{Jal01,Jac01b,Cer02,Pro02c}, and the Lyapunov regimes
\cite{Jal01,Cuc02} (for reviews see
Refs.~\cite{gor06c,jac09}). The various perturbations
considered have in common that they act ``globally'' on the
system, i.e., already a moderate perturbation strength can cause a
considerable rearrangement of the spectrum and
eigenfunctions. Correspondingly, in a semiclassical picture, a global
perturbation affects all trajectories of the system, and hence all of
them are responsible for the decay of the LE. The corresponding,
original semiclassical approach to the LE \cite{Jal01,Cuc02}, which
was recently generalized beyond the so-called diagonal approximation
\cite{gut10}, was extended in Ref.~\cite{gou07} to {\em strong local}
perturbations in coordinate space. This combined analytical
and numerical study revealed for a billiard with a local boundary
deformation, much larger than the de Broglie wavelength, a novel LE
decay law $\text{exp}(-2\gamma t)$, where $\gamma$ is the classical
``escape rate'' from the related open billiard. This approach was
refined and generalized to weak perturbations in Ref.~\cite{gou08}
predicting a nonmonotonic crossover from the FGR to the escape-rate
regime with increasing perturbation. For the case of a pistonlike
boundary perturbation the LE decay rate is expected to show distinct
oscillations as a function of the perturbation strength, i.e., piston
position. While this nonmonotonic crossover has been numerically
confirmed for maps \cite{are09}, quantum wave packet simulations
for billiards requiring more expensive numerics could only reveal
precursors of this behavior \cite{gou08}, calling for an experimental
verification of the oscillations.

For a global perturbation the fidelity decay was studied in a
microwave billiard with classically chaotic dynamics by shifting a
billiard wall \cite{sch05d}. Using the concept of scattering fidelity
\cite{sch05b} the predicted fidelity decay from the perturbative to
the FGR regime was verified experimentally.

A theoretical and experimental investigation of fidelity decay for
another type of ``local'' perturbation in the perturbative regime,
where the eigenstates are not significantly modified by the
perturbation, has been done in \cite{hoeh08a}. On the experimental
side a small scatterer was shifted inside the microwave billiard in a
two-dimensional array of pointlike scatterers. Using the random plane
wave conjecture, an algebraic decay $1/t$ was predicted theoretically
and confirmed experimentally. Another type of fidelity decay caused by
local perturbations has been studied in Ref.~\cite{koeb10},
where the coupling to an attached antenna was varied.
In the present paper we use a microwave billiard with a
piston attached to address the predicted nonmonotonic features in the
fidelity decay.

The paper is organized as follows. In Sec.~\ref{sec:Exper} we present
the experimental setup and introduce the scattering fidelity. In
Sec.~\ref{sec:theory} we briefly summarize the semiclassical results
for the LE decay with local boundary perturbations \cite{gou08} and
derive (in the Appendix) an extension of the expression for the effective
decay rate for the case of a pistonlike boundary perturbation as used
in the experiment. We then present in Sec.~\ref{sec:results} our
results for the experimentally determined scattering fidelity decay
and compare them with the theoretical predictions for the
corresponding Loschmidt echo decay. Our main findings are then
summarized in Sec.~\ref{sec:conclusions}.

\section{Experiment}
\label{sec:Exper}
Microwave experiments with flat cavities have become a well-known paradigm in the field of quantum chaos \cite{stoe99}.
In microwave billiards we can measure scattering matrix elements $S_{ab}(\nu)$ and $S_{ab}^{\prime}(\nu)$
for unperturbed and perturbed systems, independently, in frequency space.
The scattering fidelity amplitude is defined in terms of their Fourier transforms,
 $\hat{S}_{ab}$ and $\hat{S}_{ab}^{\prime}$ (upon choosing an appropriate frequency window) \cite{sch05b}:
\begin{equation}\label{eq:f_ab}
 f_{ab}(t) = \frac{\langle \hat{S}_{ab}(t)\hat{S}_{ab}^{\prime *}(t)\rangle}{ \sqrt{ \langle\hat{S}_{ab}(t)\hat{S}_{ab}^{*}(t)\rangle\langle \hat{S}_{ab}^{\prime}(t)\hat{S}_{ab}^{\prime *}(t)\rangle }}\,.
\end{equation}
The scattering fidelity itself is
\begin{equation}\label{eq:F_ab}
 F(t) = |f_{ab}(t)|^2\,.
\end{equation}
For chaotic systems and weak coupling of the measuring antenna the scattering fidelity approaches the ordinary fidelity \cite{sch05b}.

\begin{figure}[t]
\includegraphics[width=.95\columnwidth]{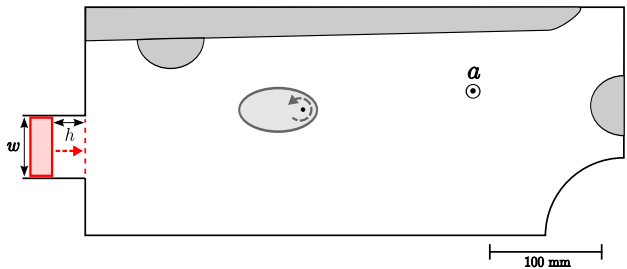}
\caption{\label{fig:01} (Color online)
Geometry of the chaotic Sinai-shaped billiard (length of 472\,mm, width of 200\,mm, and a quarter-circle of radius 70\,mm) with a variable pistonlike local boundary deformation. The piston position can be changed from a displacement $h=45$\,mm to $h=0$\,mm for four different piston widths $w$\,=\,20, 40, 70, and 98\,mm. At position $a$ the measuring antenna is introduced. The additional elements were inserted to perform ensemble averages (rotatable ellipse) and to reduce the influence of bouncing balls.}
\end{figure}

In the present experiment we chose a resonator with a height of 8\,mm
which can be considered as two-dimensional for frequencies below
18\,GHz. The setup, as illustrated in Fig.~\ref{fig:01}, is based on
a quarter Sinai-shaped billiard. Additional elements were inserted
into the billiard to reduce the influence of bouncing-ball
resonances. The classical dynamics for the chosen geometry of the
billiard is chaotic. The straight left boundary of the unperturbed
billiard was deformed at a certain position by inserting pistons of
four different widths $w$. The horizontal piston position can be
changed in steps of 0.5\,mm via a step motor from a displacement
$h=45$\,mm to $h=0$\,mm. At position $a$ an antenna is fixed and
connected to an Agilent 8720ES vector network analyzer (VNA), which
was used for measurements in a frequency range from 2 to 18\,GHz with
a resolution of 0.1\,MHz. We measured the reflection $S$-matrix
element $S_{aa}$ for four piston widths and all displacements $h$
realizing 18 different positions of a rotating ellipse (see
Fig.~\ref{fig:01}) to perform ensemble averages. The unperturbed
system is defined as the one with the straight wall, corresponding to
$h=0$\,mm.

\section{Theory}
\label{sec:theory}

The LE
\begin{equation}
\label{eq:M}	
M(t)=|\langle\phi|e^{i H^{\prime} t/\hbar}e^{-i H t/\hbar}|\phi\rangle|^2
\end{equation}
is defined as the overlap of an initial state $|\phi\rangle$ evolved
in time $t$ under a Hamiltonian $H$ with that evolved under a
perturbed Hamiltonian $H^{\prime}$. Within a semiclassical approach
this quantity was studied in Refs.~\cite{gou07,gou08} for local
perturbations in chaotic systems. There it was shown that the LE is
approximately \cite{gou08}
\begin{equation}
\label{eq:M_nd}	
 M(t)\approx e^{-\kappa\gamma t}
\end{equation}
with the effective decay rate $\kappa$ given by
\begin{equation}
\label{eq:kappa}	
\kappa=2\left(1-\mathrm{Re}\langle e^{2\pi i u/\lambda}\rangle\right) \,.
\end{equation}
Here $u$, called the deformation function, equals the length
difference, induced by the local boundary perturbation, between the
perturbed trajectory and the unperturbed one. $\lambda$ denotes the de
Broglie wavelength. For the case of a pistonlike boundary
deformation with piston width $w$ and displacement $h$ of the piston,
as it is realized in our experiment, we find, in generalization of the
results of Ref.~\cite{gou08} (see the Appendix),
\begin{equation}
\label{eq:kappa_piston_exact}	
\kappa=2-\frac{2}{w}\sum\limits_{k=0}^{\infty}\int_{\Omega_{2k+1}} \!\!\!\!\!\!\!
\textrm{d}x \, \textrm{d}\theta\cos\theta\cos\left[\frac{4\pi}{\lambda}\left(h\cos\theta+kw\sin\theta\right)\right]
\end{equation}
with the integration domains $\Omega_{2k+1}$ over incident positions
$x$ and momentum directions $\theta$ defined in
Eq.~(\ref{eq:app-4}). In the limit $h \ll w$,
Eq.~(\ref{eq:kappa_piston_exact}) reduces to \cite{gou08}
\begin{equation}
\label{eq:kappa_piston_approx}	
\kappa=\pi \mathrm{\bf H}_1(4\pi h/\lambda)
\end{equation}
with $\mathrm{\bf H}_1$ being the Struve $H$-function of first order.

Furthermore, $M(t)$ in Eq.~(\ref{eq:M_nd}) depends on $\gamma$, which is the classical escape rate of the corresponding open cavity if the piston is removed. It is given by
\begin{equation}
\label{eq:gamma}	
\gamma=\frac{p_0}{m l_d}
\end{equation}
for particles with momentum $p_0$ and mass $m$, and for the average
dwell length $l_d$ of paths in the related open chaotic billiard. In
billiards with openings (deformation widths) $w$ much smaller than the
perimeter one can approximate $l_d\approx \pi A/w$ with $A$ the area
of the corresponding closed billiard. Further we will set $p_0/m=c$,
where $c$ is, in the case of the microwave billiard the speed of light.

In semiclassical theory of the Loschmidt echo the perturbation
strength is a measure of the action change introduced by the
perturbation. Thus, for perturbations caused by pistonlike boundary
deformations the piston displacement $h$ serves as the measure of the
perturbation strength. As shown in Ref.~\cite{gou08} it is convenient
to define a dimensionless quantity
$\chi=2\pi\sqrt{\langle u^2\rangle}/\lambda$,
where $\langle u^2\rangle = 8h^2/3$ for a
pistontype deformation, as the perturbation strength. Then, based on
Eqs.~(\ref{eq:M_nd} and \ref{eq:kappa}), different decay regimes of the
LE can be identified as follows. For weak local perturbations, $\chi
\leq 1$, one has $M(t)\approx e^{-\chi^2\gamma t}$ characterizing the
FGR regime. Strong local perturbations, $\chi \gg 1$, lead to
$M(t)\approx e^{-2\gamma t}$ corresponding to the escape rate
regime. In the following section we will use Eq.~(\ref{eq:M_nd}),
together with the refined expression (\ref{eq:kappa_piston_exact}) for
the decay rate $\kappa$, for a comparison with the experimentally
determined scattering fidelity (\ref{eq:F_ab}).

\section{Results and Discussion}
\label{sec:results}

\begin{figure}
\centering
\includegraphics[width=.98\columnwidth]{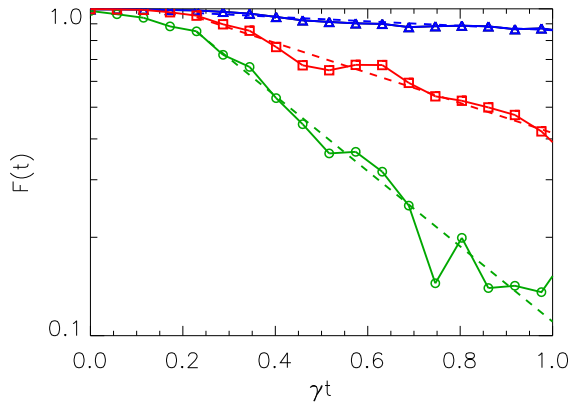}
\caption{\label{fig:02} (Color online)
Measured scattering fidelity decay $F(t)$, Eq.~(\ref{eq:F_ab}), (solid lines with symbols) for three different piston displacements $h_1=1$\,mm (blue triangles), $h_2=5$\,mm (green circles), $h_3=10$\,mm (red squares), for a frequency range $17-18$\,GHz corresponding to a mean de Broglie wavelength $\bar{\lambda}\approx 17$\,mm. The dashed lines show the corresponding semiclassical prediction, Eq.~(\ref{eq:M_nd}), for the LE decay, with $\kappa$ chosen as free parameter: $\kappa_1=0.26$, $\kappa_2=2.78$, and $\kappa_3=1.09$, respectively. The time is given in units of the dwell time $1/\gamma$, with $\gamma$ determined from experimental parameters via Eq.~(\ref{eq:gamma}) with $w=40$\,mm.
}
\end{figure}

In this section we present our measurements of the scattering fidelity decay for the pistonlike boundary
perturbation and compare them with the theoretical predictions (\ref{eq:M_nd})-(\ref{eq:kappa_piston_approx})
for LE decay for this specific type of perturbation.
We start with a piston of width $w=40$\,mm. In Fig.~\ref{fig:02} the scattering fidelity $F(t)$,
Eq.~(\ref{eq:F_ab}), is plotted for three different piston displacements $h$ acting as perturbation
to the system (symbols and solid lines).
Additionally the corresponding semiclassical predictions for the Loschmidt decay according to Eq.~(\ref{eq:M_nd}) are plotted (dashed lines) with $\kappa$ used as a fitting parameter, while $\gamma$ was obtained from the geometry.
The experimental fidelity decay shows good agreement with the expected exponential law beyond a certain time, which passes until the perturbation is ``seen'' during the measuring process. Upon increasing the displacements $h$,
illustrated in Fig.~\ref{fig:02} by the successive triangle (blue), circle (green), and squares (red) traces,
the corresponding LE decay exponent $\kappa$ exhibits a nonmonotonic behavior.

\begin{figure}
\centering
\includegraphics[width=.98\columnwidth]{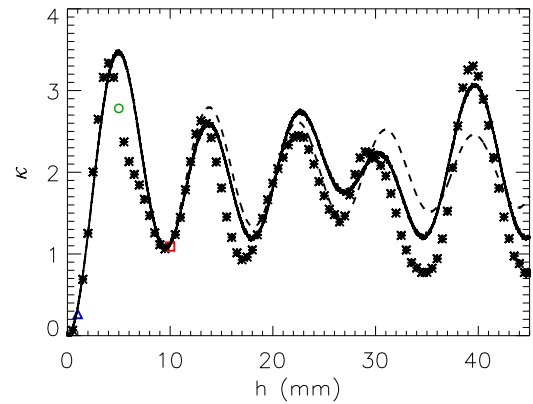}
\caption{ \label{fig:03} (Color online) Decay rate $\kappa$ as a
 function of piston displacement $h$ for a piston of width $w=40$\,mm
 in a frequency range $17-18$\,GHz corresponding to a mean de Broglie
 wavelength $\bar{\lambda}\approx 17$\,mm. The asterisks represent
 the data points obtained from fitting the decay exponent of the
 measured scattering fidelity. The three cases discussed in
 Fig.~\ref{fig:02} are marked by correspondingly colored symbols. The
 dashed curve shows the theoretical approximation
 (\ref{eq:kappa_piston_approx}) (valid for $h \ll w$), and the solid
 curve is a result of the numerical evaluation of the full
 semiclassical expression (\ref{eq:kappa_piston_exact}).}
\end{figure}

For a more detailed investigation of this dependence of the Loschmidt decay exponent $\kappa$
on the displacements $h$ of the piston, $\kappa$ is compared to the corresponding
theoretical predictions in Fig.~\ref{fig:03}.
The data points $\kappa(h)$, obtained from fitting to the experimental fidelity decay results as
in Fig.~\ref{fig:02}, are shown by asterisks. The three cases discussed in
Fig.~\ref{fig:02} are marked by correspondingly colored symbols at $h_1$ =1, 5, and 10\,mm.
The figure shows an oscillating behavior of the fidelity
exponent, and thereby of the fidelity decay at fixed time, as already predicted and referred
to as Fabry-Perot-type interferences between perturbed and unperturbed paths in Ref.~\cite{gou08}.
In Fig.~\ref{fig:03} the semiclassical results for $\kappa(h)$ resulting from the numerical evaluation of the expression
(\ref{eq:kappa_piston_exact}) and the approximation (\ref{eq:kappa_piston_approx})
(for $h \ll w$) are depicted as solid and dashed curves, respectively.
Already the dashed curve exhibits qualitatively good agreement with the measurement, though there is a
mismatch in the amplitudes: While this approximative theoretical result shows a monotonic decay of the maximum amplitude
with $h$, the experimentally observed peaks of $\kappa(h)$ do not show this simple structure. However, the refined
semiclassical prediction (\ref{eq:kappa_piston_exact}) (solid line) reflects the experimentally found irregular oscillation
of amplitudes much more convincingly, showing reasonable agreement. In particular, for $h\approx w=$\, 40\,mm
(square shape of the pistonlike deformation) the experimental results show a particularly pronounced amplitude which is met by the solid line.
In this range, which is beyond the range of validity of Eq.~(\ref{eq:kappa_piston_approx}),
the expression (\ref{eq:kappa_piston_exact}) constitutes a clear improvement.

\begin{figure}
\centering
\includegraphics[width=.98\columnwidth]{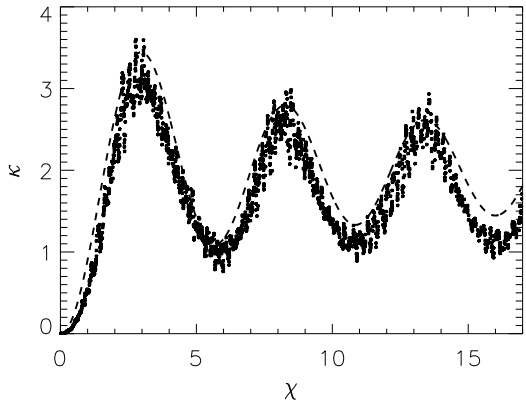}
\caption{\label{fig:04}
Decay rate $\kappa$ as a function of $\chi=\sqrt{8/3} h\,2\pi/\bar{\lambda}$. The fuzzy trace depicts
the overlayed
experimental data and the dashed curve the theoretical prediction (\ref{eq:kappa_piston_approx}).}
\end{figure}

Furthermore, in Fig.~\ref{fig:04} we present on the same plot
experimental data for $\kappa$ versus $\chi$ curves with the mean de
Broglie wavelength $\bar{\lambda}$ in the frequency range
$\bar{\lambda}<2w$, while $w\geq40$\,mm and $h<w$. We find very
convincing agreement with the theoretical prediction~(\ref{eq:kappa_piston_approx}).

\begin{figure}
\centering
\includegraphics[width=.98\columnwidth]{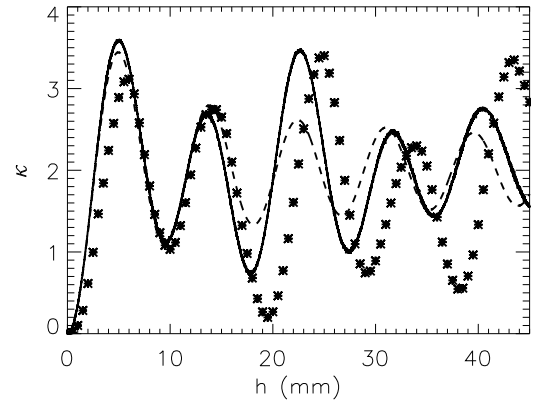}
\caption{\label{fig:05}
Decay rate $\kappa$ as a function of the displacement $h$ for a thin piston of width $w=20$\,mm
in a frequency range $17-18$\,GHz corresponding to a mean de Broglie wavelength $\bar{\lambda}\approx 17$\,mm.
The asterisks show the data points extracted again from the fit exponent of the exponential decay of
the observed scattering fidelity. The dashed and solid curves show the theoretical predictions
based on Eq.~(\ref{eq:kappa_piston_approx}) and the numerical evaluation of Eq.~(\ref{eq:kappa_piston_exact}).}
\end{figure}

Finally, we demonstrate that the agreement between the experimental
and theoretical curves can be shaken by pushing the experimental
conditions too far beyond the main limit of the semiclassical theory,
$\lambda \ll w$. Figure~\ref{fig:05} shows the decay rate $\kappa$
for a piston width $w=20$\,mm which is of the order of
$\bar{\lambda}$. As expected, the agreement between theory and
experiment is not as good as that for the $w=40$\,mm case; in particular the
experimental data points (dashed line) oscillate with a period that
differs from the theoretical one. Experimentally, we again find a
more pronounced amplitude around $h\approx w$, which is again
described more convincingly by the full (solid line) than the
approximative theoretical expression. However, the fact that the
experimental parameters are beyond the regime of validity of the
semiclassical theory does not allow for a further reasonable
comparison between experiment and theory.

\section{Conclusions}
\label{sec:conclusions}

In this work we presented an experimental verification of the
recent semiclassical predictions for fidelity decay arising from
a local perturbation of a chaotic quantum system. In particular, we
could confirm that the rate governing exponential fidelity decay
exhibits oscillations as a function of the perturbation strength. The
observed nonmonotonic behavior implies that for certain ranges of the
perturbation strength the fidelity decay becomes weaker (for fixed
time) with increasing perturbation strength. While the original
semiclassical treatment \cite{gou08} for a pistontype local boundary
deformation was based on the assumption of a small piston
displacement, the present microwave setting required a generalization
of the semiclassical approach to arbitrary ratios between piston
displacement and width, which we performed by deriving an expression
for the decay exponent in terms of a quadrature. We find quantitative
agreement between the measurements and this refined semiclassical
theory despite the fact that the microwave billiard does not really
satisfy the underlying semiclassical assumption, namely, that the
extent of the local perturbation, here the piston width $w$, should be
much larger than the de~Broglie wave length $\lambda$. An improved
semiclassical approach for local perturbations of size $w \lesssim
\lambda$ would require one to treat semiclassical contributions due to
diffractive trajectories properly, which is left for future research.

On the experimental side, there remains the challenge to observe
fidelity decay in the escape rate regime (for strong perturbations)
characterized by a perturbation-independent fidelity decay
rate. Naturally, this regime is difficult to access since the expected
signals are tiny.

\begin{acknowledgments}
 We thank Thomas Seligman for helpful discussions at an early stage
 which partly triggered this work. We are also thankful to Rodolfo
 Jalabert for helpful correspondence. B.K., U.K., H.-J.S., and K.R.\ acknowledge
 funding from the Deutsche Forschungsgemeinschaft through the
 research group FOR760 ``Scattering Systems with Complex
 Dynamics''. A.G.\ acknowledges the support by EPSRC under Grant
 No.~EP/E024629/1.
\end{acknowledgments}

\appendix

\section*{Appendix: Semiclassical theory of exponential fidelity decay for
 arbitrary $w$ and $h$}
\label{app:Iwh}

\begin{figure}[t]
\includegraphics[width=.95\columnwidth]{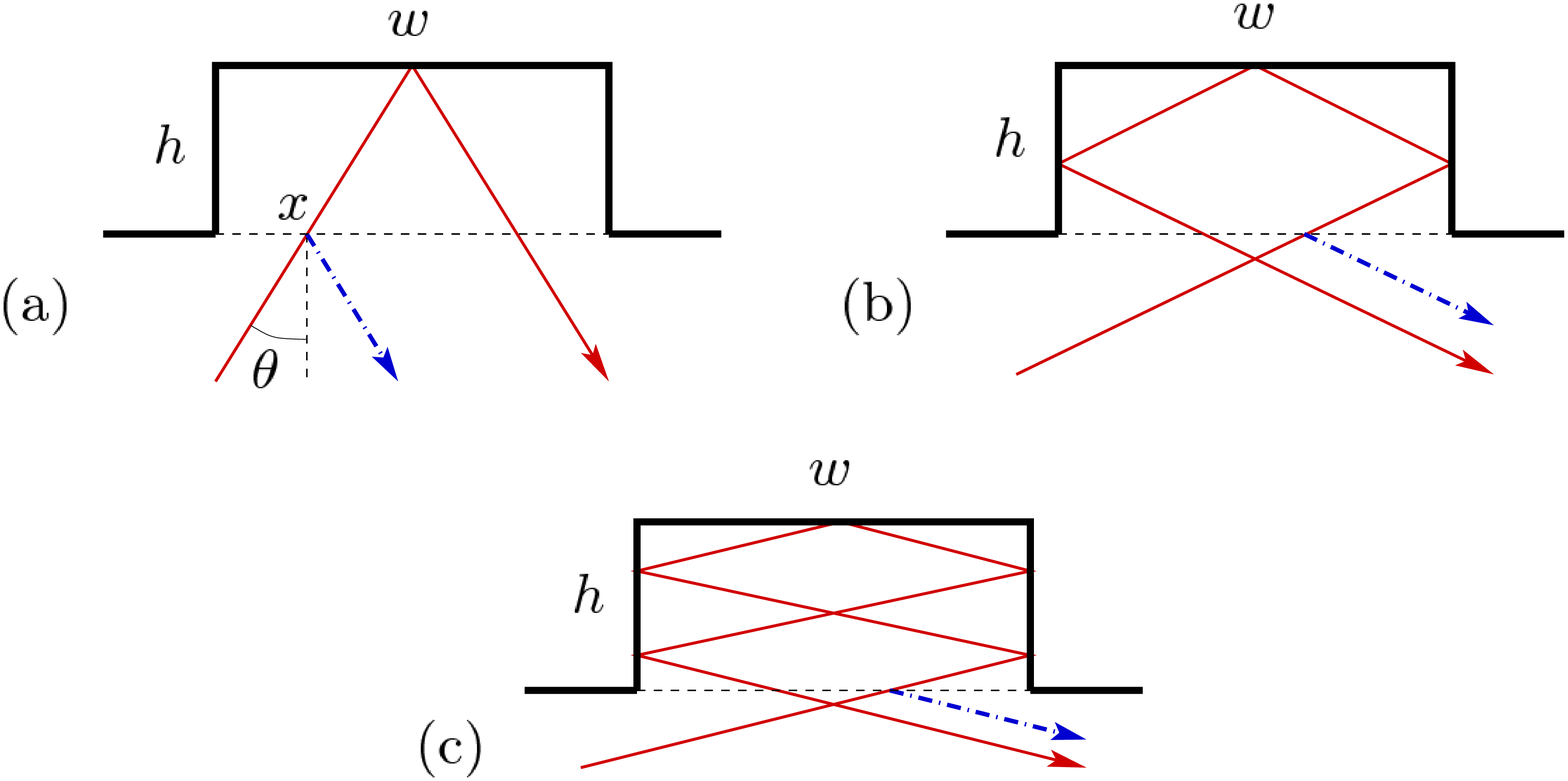}
\caption{\label{fig:06} (Color online) Examples of correlated
 trajectory pairs, unperturbed (blue dash-dotted line) and perturbed
 (red solid line), belonging to sets $\Omega_1$ (a), $\Omega_3$ (b),
 and $\Omega_5$ (c) (see text in the Appendix).}
\end{figure}

In view of the above experiments we extend the semiclassical theory for the decay
of the fidelity due to pistonlike boundary deformations, presented in
Ref.~\cite{gou08} for the limit $h \ll w$. There it was shown that the rate of exponential
fidelity decay is given by Eq.~(\ref{eq:kappa}) with
\begin{equation}
 \langle e^{2\pi i u/\lambda}\rangle = \int_0^w \frac{\mathrm{d}x}{w}
 \int_0^{\pi/2} \!\! \mathrm{d}\theta \cos\theta
 \, e^{2\pi i u(x,\theta)/\lambda} \,.
\label{eq:app-1}
\end{equation}
Here $x$ and $\theta$ denote the incident position and angle, respectively
[see Fig.~\ref{fig:06}(a)]. The deformation function $u(x,\theta)$
equals the length difference between the perturbed and unperturbed
trajectory of a correlated trajectory pair. A pair made up of an unperturbed and
perturbed trajectory is considered correlated if the two
trajectories exit the perturbation region with the same momentum direction \cite{gou08}
(see Fig.~\ref{fig:06}).

We first note that only perturbed trajectories with an odd number of
reflections may exit the perturbation region with the same momentum
direction as the unperturbed trajectory and, therefore,
contribute to the fidelity. We denote by $\Omega_n$ a set of all correlated
trajectory pairs where the perturbed trajectory exhibits $n$
reflections. The panels (a)-(c) in Fig.~\ref{fig:06} show representative
trajectory pairs belonging to the sets $\Omega_1$, $\Omega_3$ and
$\Omega_5$, respectively. Equation~(\ref{eq:app-1}) can then
be written as
\begin{equation}
 \langle e^{2\pi i u/\lambda}\rangle = \frac{1}{w} \sum_{k=0}^\infty
 \int_{\Omega_{2k+1}} \!\!\!\!\!\!\! \mathrm{d}x \, \mathrm{d}\theta \cos\theta \, e^{iu/\lambda} \,,
\label{eq:app-2}
\end{equation}
where the double integral in the $k$th summand runs over a region in
the $(x,\theta)$ plane that defines the set $\Omega_{2k+1}$.

\begin{figure}[t]
\includegraphics[width=.95\columnwidth]{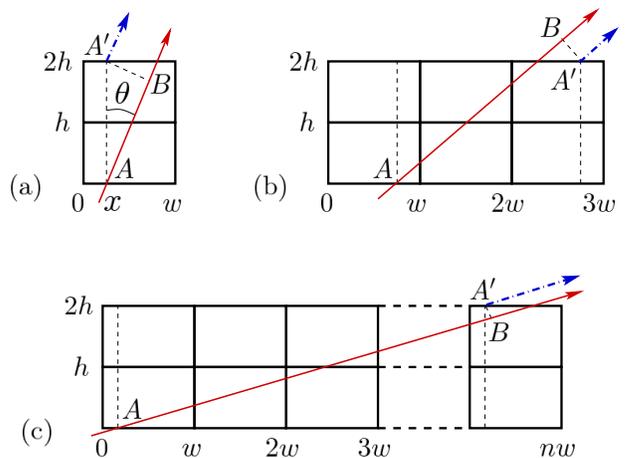}
\caption{\label{fig:07}(Color online)
 ``Unfolded'' representation of correlated
 trajectory pairs belonging to sets $\Omega_1$ (a), $\Omega_3$ (b),
 and $\Omega_n$ (c) with an odd integer $n$.}
\end{figure}

In order to calculate the deformation function $u(x,\theta)$ for a
trajectory pair from the set $\Omega_n$ we ``unfold'' the perturbation
rectangle (by ``gluing'' mirror copies of the rectangle along the
reflection sides) making the perturbed trajectory become a straight
line (see Fig.~\ref{fig:07}). Thereby, Fig.~\ref{fig:07}(a) is the
``unfolded'' version of Fig.~\ref{fig:06}(a), Fig.~\ref{fig:07}(b)
corresponds to Fig.~\ref{fig:06}(b), and Fig.~\ref{fig:07}(c)
represents a trajectory pair belonging to $\Omega_n$. In this representation, the
deformation function $u$ equals the length of the interval $AB$, $u =
\overline{AB}$. Here $A$ is a point of incidence, while the point $B$ belongs
to the perturbed trajectory and is specified by requiring the angle
$\widehat{ABA'}$ to be $\pi/2$, where $A'$ represents the incident
point $A$ in the ``exit'' copy of the perturbation rectangle (see
Fig.~\ref{fig:07}). Then, a geometrical calculation yields
\begin{equation}
 u(x,\theta) = 2h \cos\theta + (n-1)w \sin\theta \,,
\label{eq:app-3}
\end{equation}
where $(x,\theta) \in \Omega_n$.

\begin{figure}[t]
\includegraphics[width=.95\columnwidth]{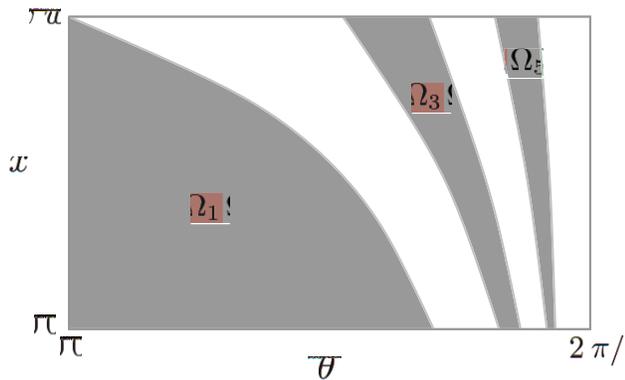}
\caption{\label{fig:08} Schematic representation of
 the regions $\Omega_1$, $\Omega_3$, and $\Omega_5$, see
 Eq.~(\ref{eq:app-4}). Further regions, $\Omega_{2k+1}$ with $k\ge3$,
 contributing to the sum on the right hand side of
 Eq.~(\ref{eq:kappa_piston_exact}) are not shown in the figure; they
 cluster as narrow stripes ``to the right'' of $\Omega_5$ and
 approach $\theta = \pi/2$ in the limit $k \rightarrow \infty$.}
\end{figure}

We now give a precise definition of the region $\Omega_n$ in the
$(x,\theta)$ plane. As evident from Fig.~\ref{fig:07}(c), a trajectory
pair belongs to $\Omega_n$ if and only if the $x$ coordinate of the
exit point of the perturbed trajectory (in the unfolded picture)
lies between $(n-1)w$ and $nw$. This yields
\begin{align}
 \Omega_n = \Big\{ (x,\theta) \, : \quad &x\in(0,w)\,, \;
 \theta\in(0,\pi/2)\,, \nonumber\\ &x+2h\tan\theta \in (n-1,n)w
 \Big\} \,.
\label{eq:app-4}
\end{align}
Figure~\ref{fig:08} schematically shows the first three sets, $\Omega_1$, $\Omega_3$, and $\Omega_5$, contributing to the sum in Eq.~(\ref{eq:app-2}).

Combining Eqs.~(\ref{eq:kappa}), (\ref{eq:app-2}), (\ref{eq:app-3}), and (\ref{eq:app-4}) we arrive at Eq.~(\ref{eq:kappa_piston_exact}), which completes the derivation. The sum and integrals in Eq.~(\ref{eq:kappa_piston_exact}) are then computed numerically by means of Monte Carlo sampling.

As a final remark, we note that in the limit of $h \ll w$ the sum on the right hand side of Eq.~(\ref{eq:kappa_piston_exact}) is dominated by the $k=0$ term. Then the integration region $\Omega_1$ can be approximately extended to the rectangle $x \in (0,w)$, $\theta \in (0,\pi/2)$, since the contribution for large angles, $\theta$ close to $\pi/2$, is suppressed by the $\cos\theta$ term in the integrand. This approximation leads to Eq.~(\ref{eq:kappa_piston_approx}) (see also Ref.~\cite{gou08}).

%%%%%%%%%%%%%%%%%%%%%%%%%%%%%%%%%%%%%%%%%%%%%%%%%%%%%%%%%%%%%%%%%%%%%%%
%merlin.mbs 2010-03-15 4.21a (PWD, AO, DPC)
%Control: key (0)
%Control: author (8) initials jnrlst
%Control: editor formatted (1) identically to author
%Control: production of article title (-1) disabled
%Control: page (0) single
%Control: year (1) truncated
%Control: production of eprint (0) enabled
%
\end{document}